\documentclass[conference, 10pt]{IEEEtran}

\usepackage[latin9]{inputenc}
\usepackage{graphicx}
\usepackage{listings}
\usepackage[table]{xcolor}  
\usepackage{flushend}

\usepackage{cite}
\usepackage{url}				 

\usepackage{amssymb}				



\begin{document}


\IEEEoverridecommandlockouts

\title{\huge Beamformed Fingerprint Learning for \\ Accurate Millimeter Wave Positioning}

%
%

\author{\IEEEauthorblockN{Jo\~{a}o Gante\IEEEauthorrefmark{1}
Gabriel Falc\~{a}o\IEEEauthorrefmark{2}
Leonel Sousa\IEEEauthorrefmark{1}}
\IEEEauthorblockA{\IEEEauthorrefmark{1}INESC-ID, IST, Universidade de Lisboa, Portugal}
\IEEEauthorblockA{\IEEEauthorrefmark{2}Instituto de Telecomunica\c{c}\~{o}es, Department of Electrical and Computer Engineering, University of Coimbra, Portugal}
\thanks{This work was supported by national funds through Funda\c{c}\~{a}o para a Ci\^{e}ncia e a Tecnologia (FCT) with reference UID/CEC/50021/2013, and FCT Grant No. FRH/BD/103960/2014.}}

\IEEEaftertitletext{\vspace{-7mm}}

\maketitle

\begin{abstract}

With millimeter wave wireless communications, the resulting radiation reflects on most visible objects, creating rich multipath environments, namely in urban scenarios. The radiation captured by a listening device is thus shaped by the obstacles encountered, which carry latent information regarding their relative positions. 

In this paper, a system to convert the received millimeter wave radiation into the device's position is proposed, making use of the aforementioned hidden information. Using deep learning techniques and a pre-established codebook of beamforming patterns transmitted by a base station, the simulations show that average estimation errors below $\mathbf{10}$ meters are achievable in realistic outdoors scenarios that contain mostly non-line-of-sight positions, paving the way for new positioning systems.

\end{abstract}

\begin{IEEEkeywords} 5G, Beamforming, Deep Learning, mmWaves, Outdoor Positioning.\end{IEEEkeywords}

\IEEEpeerreviewmaketitle{}

\section{Introduction}


Through 5G related research, the door to the so called millimeter wave (mmWave) frequencies reopened, unlocking a huge chunk of untapped bandwidth\cite{5g}. With mmWaves, the propagation changes dramatically: the resulting radiation has severe path loss properties and reflects on most visible obstacles \cite{mmwaves_2}. To counteract the aforementioned characteristics, beamforming (BF) is usually employed in systems containing multiple-input and multiple-output (MIMO) antennas, enabling steerable and focused radiation patterns. 

With that recent focus on mmWaves, new positioning systems based on these frequencies were proposed \cite{network_location_survey}. The achievable accuracy in controlled conditions is remarkable, with sub-meter accuracy in indoor \cite{indoor_positioning_1m} and ultra-dense line-of-sight (LOS) outdoor scenarios \cite{outdoor_los_1m}. Nevertheless, in order to be useful in outdoor scenarios, a mmWave positioning system must also be able to deal with devices in non-line-of-sight (NLOS) locations. 

The works developed in \cite{mmwave_location_cs, mmwave_location_lf, mmwave_location_iter, outdoor_fingerprint_35m} attempt to address this concern, being capable of locating devices in both LOS and NLOS situations. The method in \cite{mmwave_location_cs} applies compressed sensing on information gathered from static listeners, while in \cite{mmwave_location_lf} multiple access points are used to create a location fingerprint database of received powers and angles-of-arrival (AoA). In \cite{mmwave_location_iter}, the authors use multiple BF transmissions and an iterative algorithm to estimate the position and orientation of the device. Nevertheless, the methods proposed in \cite{mmwave_location_cs, mmwave_location_lf, mmwave_location_iter} have difficulties complying with typical outdoor situations: \cite{mmwave_location_cs} and \cite{mmwave_location_lf} assume that each device is always in range of multiple static transceivers, while \cite{mmwave_location_iter} struggles with NLOS locations, requiring multiple transmission paths reflecting in at least three different surfaces. Finally, the method proposed in \cite{outdoor_fingerprint_35m} overcomes the aforementioned restrictions by creating a fingerprint database of uplink pilots transmitted to a distributed massive MIMO base station (BS), and then resolving the position using a Gaussian process regression, obtaining a root-mean-square-error (RMSE) of $35$m.

\begin{figure}[t!]
  \centering
	\includegraphics[width=82mm, trim={6mm 7mm 6mm 5mm}]{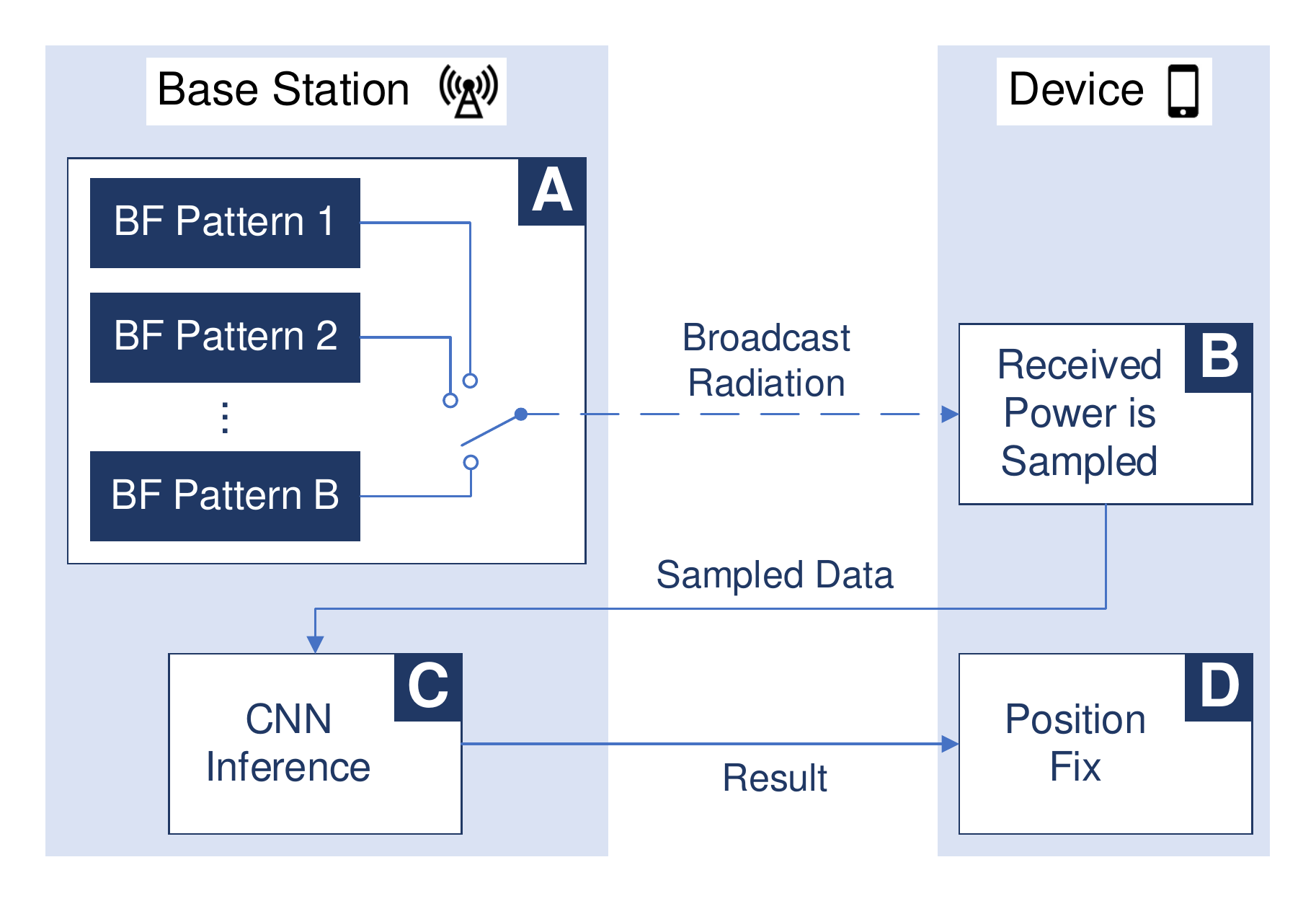}
	\caption{Overall scheme of the proposed system. The device samples the received power from radiation transmitted using a fixed set of beamforming patterns, resulting in a unique data arrangement that can then be translated into its position.}
	\label{fig:diagrama1}
	\vspace{-4mm}
\end{figure}

For a mobile mmWave receiver connected to a BS, the received power delay profile (PDP) for a given BF configuration is determined by the receiver's location. Therefore, if a BS transmits short pulses employing a sequence of directive BF patterns, so as to cover all possible angles of transmission, each position covered by the system will likely have a unique pattern in terms of received power over time. In this paper, we discuss how to obtain said patterns and propose to apply convolutional neural networks (CNN) so as to learn them, providing geolocation to devices with mmWave capabilities. 

By piggybacking on the (planned) infrastructure and hardware, the implementation cost can be kept low. To simulate the feasibility of the proposed system, a propagation dataset is generated using ray-tracing simulations on accurate 3D maps. Ultimately, by being able to provide accurate positioning in outdoors NLOS scenarios, this technique can be used to complement the existing mmWave positioning techniques, resulting in an enhanced experience for the user.

\section{Proposed System Description\label{system}}


With the use of mmWaves on MIMO systems, the resulting multipath propagation is defined by the used BF and the existing obstacles. For 5G BSs, which are expected to be located in elevated positions of urban scenarios, most of the obstacles are static for a significant amount of time, as they are predominantly buildings. Therefore, successive measurements of the received radiation pattern at a given position are expected to remain comparable until a significant change in the surrounding area occurs. 

In order to infer the position from the hypothetical information contained in the received radiation, two requirements have to be respected: \textit{i)} the adopted transmitter BFs must be constant, to culminate in an equivalent set of transmitted radiation patterns, and \textit{ii)} the target devices must be able to detect the transmitted radiation with the same detection scheme, to gather the required information.


To comply with both requirements, the system depicted in Fig. \ref{fig:diagrama1} is proposed. It contains four distinct phases, as labeled in the diagram, whose details are described bellow. Phase A will broadcast pulse waveforms using constant set of radiation patterns, while phase B focuses on measuring the resulting radiation at the target device. After all the required measurements are performed and transmitted to the BS, phase C infers the device's position, which will be relayed back to the device in phase D. 

\textbf{Phase A}: Since mmWave transmissions must employ directive (and therefore, narrow) beam patterns, a fixed codebook with $B_{Tx}$ BFs is proposed, so as to fully cover all possible angles of transmission. Assuming a BS with $N_S$ antennas, the frequency-domain signal at the $N_R$ mobile devices antennas $\mathbf{y}\in\mathbb{R}^{N_R \times 1}$ can be written as
%
\begin{equation}
\mathbf{y} = \mathbf{Hf} x + \mathbf{z},
\label{eq:signal_at_receiver}
\end{equation}   
where $\mathbf{H}\in\mathbb{C}^{N_R \times N_S}$ is the channel matrix, $\mathbf{f}\in\mathbb{C}^{N_S \times 1}$ denotes the currently selected beamforming, $x\in\mathbb{C}$ is the waveform to be detected, and $\mathbf{z}\in\mathbb{C}^{N_R \times 1}$ is the noise.

In order to avoid losing information due to destructive interference, the transmissions using those codebook entries should have a minor time interval between them ($T_{guard}$), to account for longer paths with multiple reflections.

\textbf{Phase B}: To capture the detail inherent to the radiation power over time patterns, the system must be able to measure the transmitted pulses at a high rate. As the data patterns must be consistent regardless of the target device, the sampling rate must be equal for all the receivers, and those receivers must be synchronized with the BS. Furthermore, if BF at the receiver is also included in the system, a specific array gain must also be set for all receivers (to ensure consistency). In that case, the receiver would also have to define a BF codebook to search over all AoAs with similar gain, containing $B_{Rx}$ entries, and it would have to sample the original transmission $B_{Rx}$ times, storing the maximum measured value for each data point. The acquired data from transmit BF $b$, $\mathbf{d}_b$, can thus be written as
\begin{equation}
d_b[n] = \max\limits_{i = 1,\dots, B_{Rx}} y_i(nT), \quad n = 0, 1,\dots, N-1,
\label{eq:digital_signal}
\end{equation}  
where $T$ is the sampling period and $N$ is the number of samples to gather per BF. To be effective, $T$ (and the pulse duration) must be smaller than $100$ ns, as it will examined in the following section. Nowadays smartphones can connect to LTE networks with bandwidths exceeding $10$ MHz, so, in order to use the proposed system, no additional hardware should be required at a device with mmWave capabilities.

\begin{figure}[t!]
  \centering
	\includegraphics[width=88mm, trim={8.8mm 0mm 2mm 0mm}]{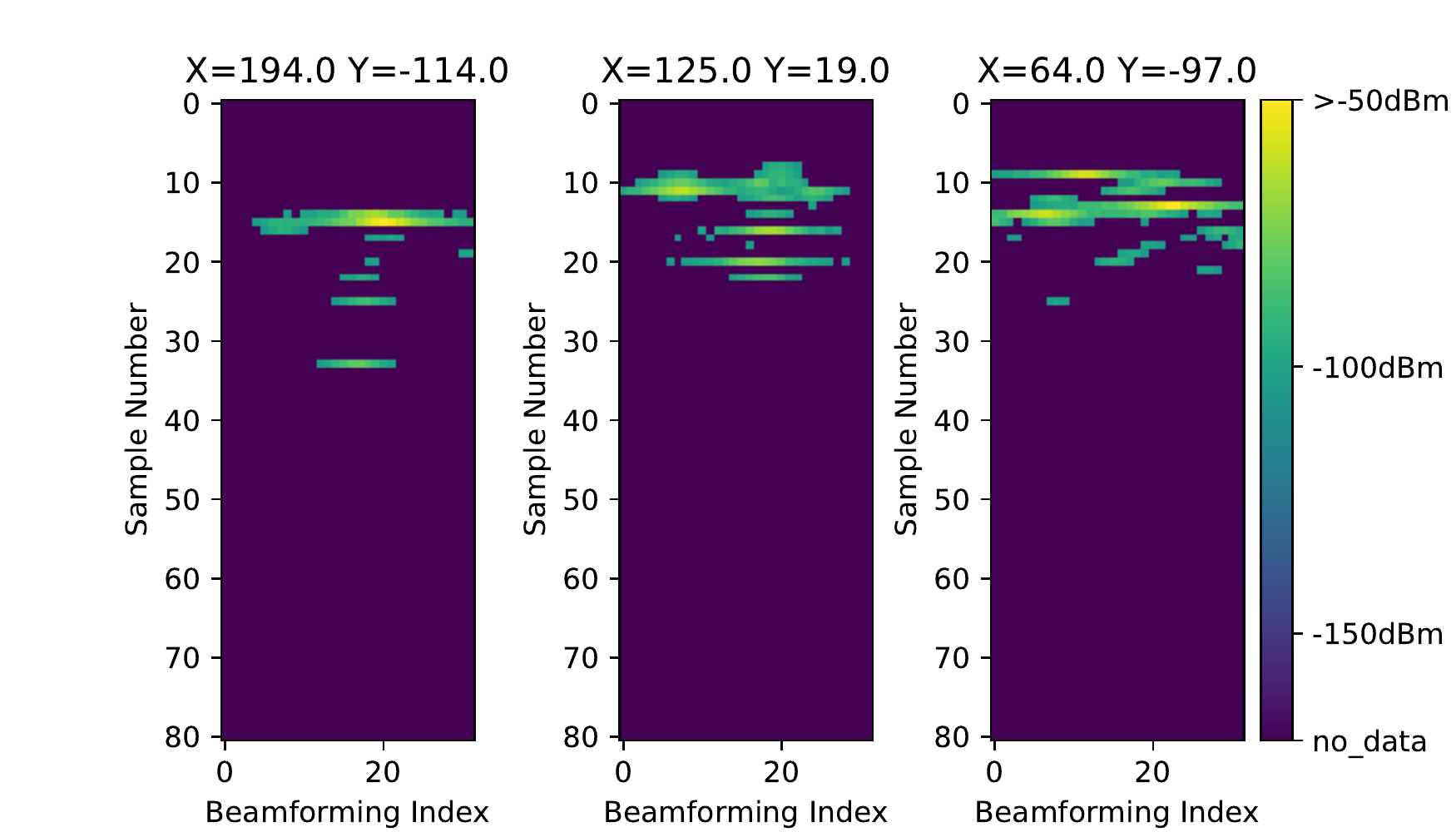}
	\caption{Noiseless data samples from the experimental simulations, before removing low-power entries due to the detection threshold. Adjacent beamforming patterns have a high probability of having similar patterns.}
	\label{fig:data_examples}
\end{figure}

\textbf{Phases C and D}: Upon obtaining the required data from all the BFs in the codebook ($\mathbf{d}$), performing inference on a trained CNN will result in a position estimation. The proposed system can thus be considered fingerprint-based, since the CNN must be trained with data from the possible positions. The inference can either be performed on the device, or at the base station (after transmitting the required data from the device, and then reporting the result back to it). However, performing the inference at the device has its limitations, since a CNN can easily have millions of weights. In that case, the device would have to download (and store) all those weights each time it connects to a new BS, or when a network is retrained. For those reasons, we propose to perform the inference at the BS, as depicted in Fig. \ref{fig:diagrama1}

\begin{figure}[t!]
  \centering
	\includegraphics[width=88mm, trim={4mm 0mm 4mm 0mm}]{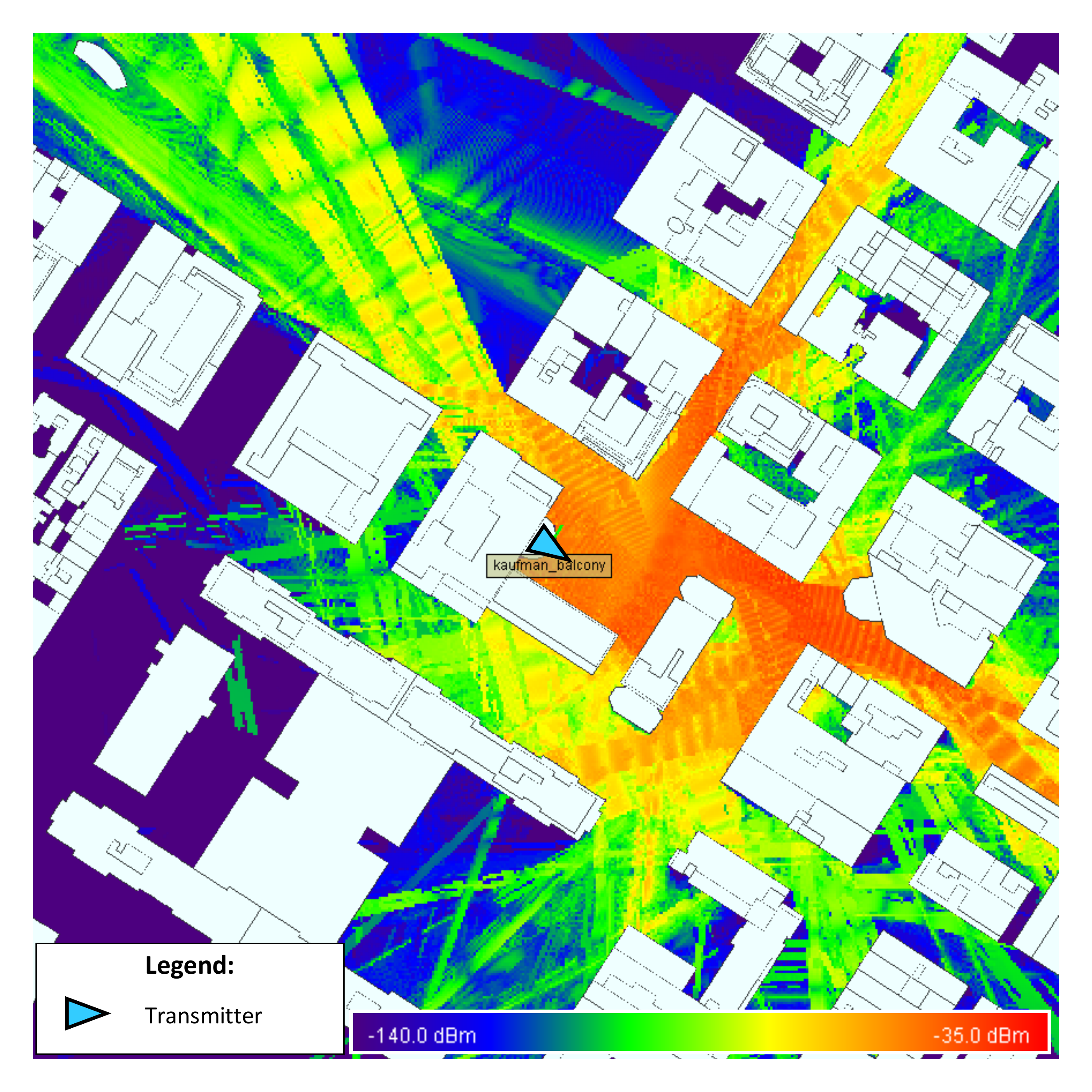}
	\caption{Ray-tracing simulation in the NYU area, using the parameters in table \ref{tab:ray_spec} with a transmit power of $30$ dBm. The results shown correspond to the \textit{maximum received power for all possible transmit BFs}. In \cite{meu_icassp_2018}, it was shown that this simulation matched the experimental measurements in \cite{mmwave_measurements}.}
	\label{fig:nyu_sim}
\end{figure}

\section{Data Analysis\label{learning_data}}

The accuracy resulting from the proposed system ultimately depends on the learning capabilities of the inference block. The system performance is thus determined by two factors: the data obtained in phase B and the learning method used in phase C.

As mentioned in section \ref{system}, the used pulse duration in phase A and the measurement mechanism selected for phase B should be able to provide a temporal resolution exceeding $100$ ns. In such conditions, the received radiations will arrive in clusters with voids large enough to be reliably detected \cite{mmwave_measurements}, and the resulting data will be sparse (as it is observable in Fig. \ref{fig:data_examples}). The ability to distinguish these voids provides meaningful shape to the resulting data, enhancing the learning capabilities. In fact, as it will be observable in section V, the position of the acquired non-zero samples in the data yields more information than their magnitude, especially in noisy environments (where the fluctuations are more likely to be visible at the amplitude level). Therefore, a binary detection of the signal's existence can be used, further reducing the requirements for the proposed system. 


When visualizing the sampled data, it is possible to represent it in a three dimensional space: time, power and  BS BF index. It is interesting to notice a visual pattern in the data, when the sequence of BF indexes correspond to a continuous sweep over the azimuth. If that resulting data is plotted as a 2D image, where the axis are the time and the BF index, the formed image will likely have short lines over the BF dimension, as seen in the examples in Fig. \ref{fig:data_examples}. In other words, this means that physically adjacent BF patterns will likely have similar clusters to the same location, and thus carry some redundant information. As result, adding an increasing number of BF patterns to phase A will have diminishing returns on the position inference accuracy.

The inference in the proposed system can be classified as a regression problem, since it will be transforming data to a continuous set of results (\textit{i.e.}, the device position). Due to performance, suitability, and noise-resisting capabilities for pattern recognition problems, CNNs were selected as learning mechanism \cite{dnn_nature}. Using a CNN, the system should be able to cope with the non-linearities introduced by reflections and other propagation artifacts. For the CNN cost function, the minimum mean square error (MMSE) was selected, \textit{i.e.},
\begin{equation}
\mathbf{\hat{p}_{MMSE}\left(d\right)} = argmin_\mathbf{{\hat{p}}}\ \mathbf{E\{ \left(\hat{p} - p \right)^T \left(\hat{p} - p \right) \}},
\label{eq:mmse}
\end{equation} 
where $\mathbf{\hat{p}}$ denotes the estimated position given the input data $\mathbf{d}$. The usage of this cost function can be interpreted as a minimization of the squared distance between the real position $\mathbf{p}$ and its estimate.

\begin{table}[t!]
	\centering
	\caption{Ray-Tracing Simulation Parameters}
	\label{tab:ray_spec}
	\begin{tabular}{lc}
		\hline
		\rowcolor[HTML]{C0C0C0}
		\textbf{Parameter Name}        & \textbf{Value}    \\
		\hline
		Carrier Frequency & 28 GHz   \\
		Transmit Power & 45 dBm   \\
		Tx. Antenna Gain & 24.5 dBi (horn antenna) \\
		HPBW & 10.9$^\circ$ \\
		Transmitter Downtilt & 10$^\circ$ \\
		Codebook Size & $32$ ($155^\circ$ arc with $5^\circ$ between entries) \\
		Receiver Grid Size & $160801$ ($400\times400$ m, $1$ m between Rx, \\
		 & $1$ m above the ground) \\
		\hline
	\end{tabular}
\end{table}

\begin{table}[t!]
	\centering
	\caption{Convolutional Neural Network and Data Parameters}
	\label{tab:cnn_spec}
	\begin{tabular}{lc}
		\hline
		\rowcolor[HTML]{C0C0C0}
		\textbf{Parameter Name}        & \textbf{Value}    \\
		\hline
		Convolutional Layers (CL) & $1$ ($8\ 1\times3$ filters,\\
		& followed by $2\times1$ max-pooling)  \\
		Hidden Layers (HL) & $7$ ($1024$ neurons each)  \\
		HL Activation Function & Rectified Linear Unit \\
		Output Layer & $2$ Linear Neurons (2D position)\\
		Optimizer Function & ADAM \\
		Learning Rate & $10^{-5}$ ($\times 0.99$ decay each epoch) \\
		Training Epochs & $1000$ \\
		Dropout Rate & $0.5$ \\
		\hline
		Samples per Tx. BF & $82$ ($4.1\ \mu$s @ $20$ MHz) \\
		Assumed Rx. Gain & $10$ dBi\\
		Detection Threshold & $-100$ dBm \\
		Data Preprocessing & Row Normalization / Binarization \\
		Added Noise & $\sigma$ = $\left[0,10\right]$ dB (Log-Normal) \\
		\hline
	\end{tabular}
\end{table}

\section{Simulation Apparatus}

To simulate the proposed system accuracy, a dataset using mmWave ray-tracing simulations in the NYU area was created, containing fingerprint data from $160801$ positions. These ray-tracing simulations were created with the Wireless InSite 3.0.0.1 ray-tracing software \cite{wireless_insite}, using the high precision open-source 3D map made available by the New York City Department of Information Technology \& Telecommunications \cite{nyc_3d_map}, as well as the specifications depicted in table \ref{tab:ray_spec} (mostly inherited from \cite{mmwave_measurements}). In \cite{meu_icassp_2018}, it was shown that the ray-tracing simulations (presented in Fig. \ref{fig:nyu_sim}) matched the experimental measurements obtained in \cite{mmwave_measurements}.

Even though actual BF is not performed in the ray-tracing simulations (due to software inability), a physically rotating horn antenna produces similar directive radiation patterns. The received power over time data was sampled at $20$MHz over $4.1\ \mu$s (which contained $99$\% of the non-zero entries). Combined with the $32$ transmit directions, this resulted in $2624$ floating-point elements per location entry, as represented in Fig. \ref{fig:data_examples}. Regarding the BF at the receiver, a $10$ dBi gain was considered, akin to the mmWave mobile device tested in \cite{samsung_2016_2}, which is able to cover most of the receive directions employing a codebook with $16$ elements. The resulting data was labeled with the corresponding bidimensional position, in a area centered at the base station.

\begin{figure}[t!]
  \centering
	\includegraphics[width=90mm, trim={4mm 5mm 2mm 5mm}]{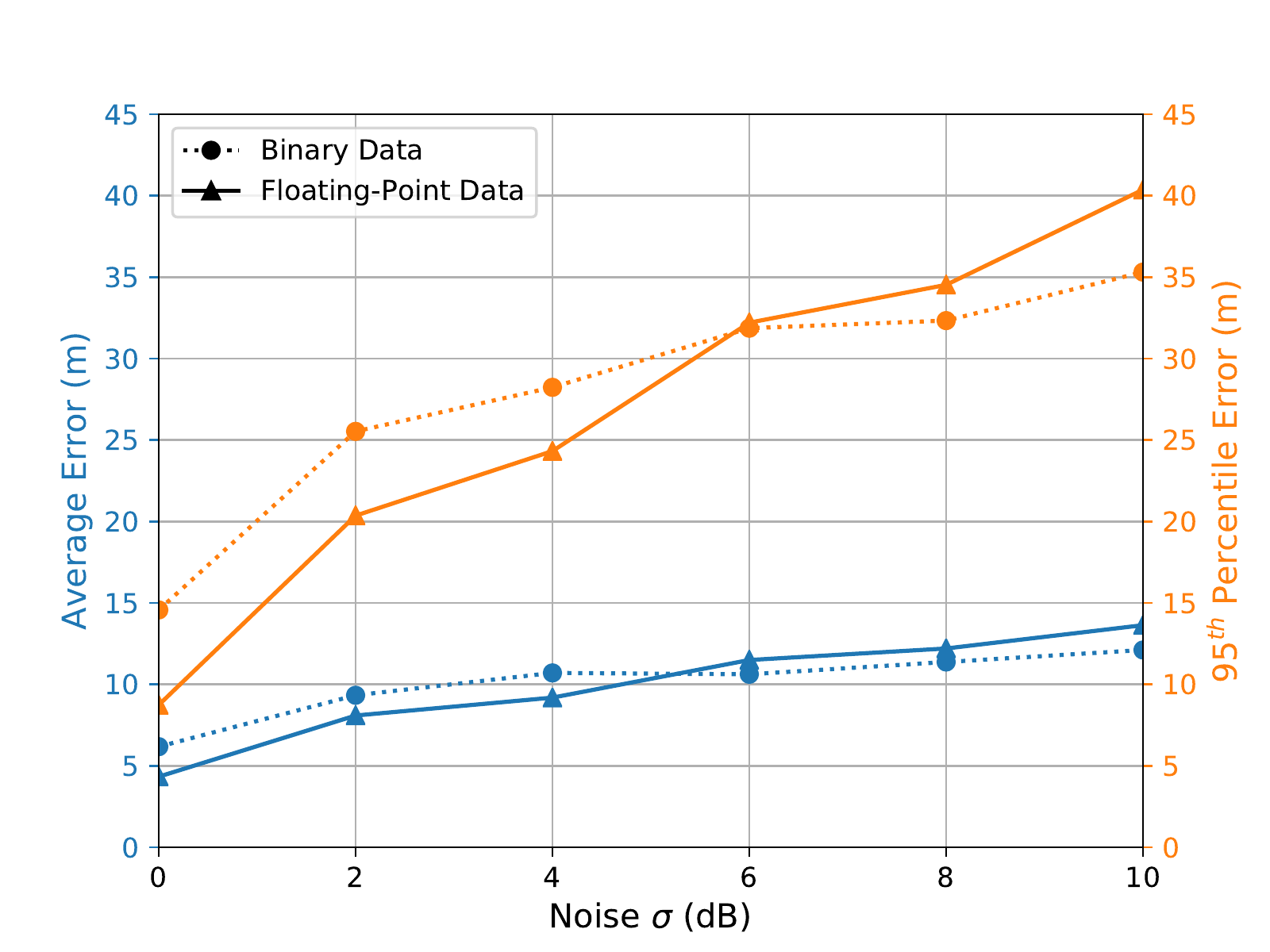}
	\caption{Average and $95^{th}$ percentile prediction errors for the proposed system, for multiple log-normal modeled noise levels, depicted by its standard deviation ($\sigma$). For larger noise values, using binary data resulted in improved performance.}
	\label{fig:noise_vs_error}
\end{figure}

When noise is considered in the proposed system, it is added to the ray-tracing data following a log-normal distribution. This happens before discarding any data entry below the detection threshold of $-100$ dBm (selected due to the thermal noise for the considered bandwidth). If the simulated system considers a binary detection of the received signal, the simulation data is transformed after adding the noise and passing through the detection threshold. This data is then sampled for each position that contains at least one non-zero entry, so as to create a training set (for each CNN training epoch, a new training set is generated). Since the system is expected to be used to predict positions for which it already has samples, the test set is generated as the training set. When evaluating the average results, a total of $10$ test sets are used.

The selected CNN follows a traditional architecture, whose hyperparameters were selected after empirical tests, also taking in account the total execution time (the training takes less than $8$ hours in a Nvidia GTX 780 Ti, using Google's TensorFlow framework). The system can thus be summarized as depicted in table \ref{tab:cnn_spec}\footnote{The simulation code and used data are available on https://github.com/gante/mmWave-localization-learning}.

\begin{figure}[t!]
  \centering
	\includegraphics[width=90mm, trim={4mm 5mm 2mm 5mm}]{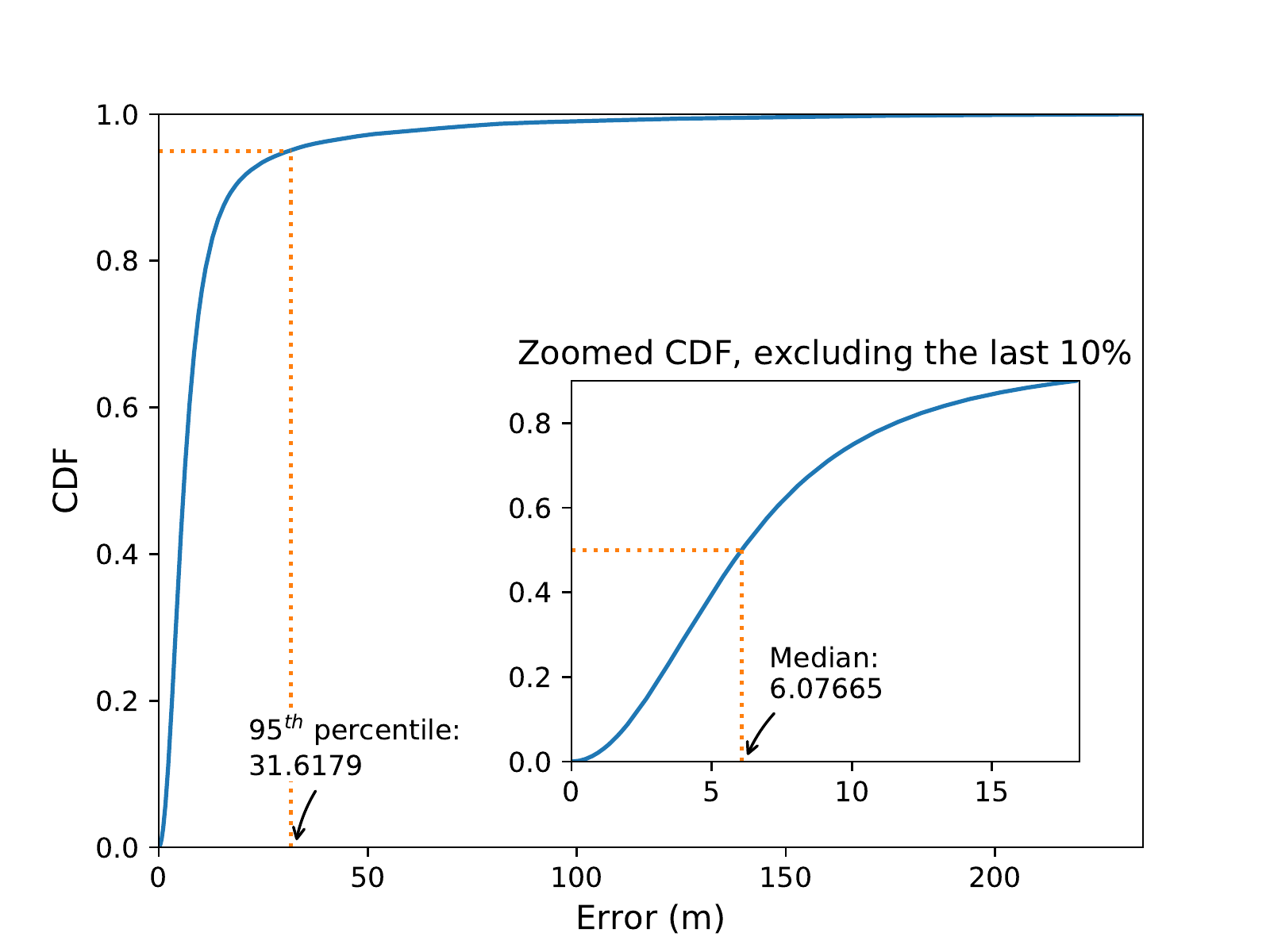}
	\caption{CDF of the prediction errors, considering binary data samples and a noise $\sigma$ of $6$ dB. The least accurate predictions are very inaccurate, and thus the average error is larger than the median error.}
	\label{fig:cdf}
\end{figure}

\section{Simulation Results}

When working with the proposed system, the data format must be carefully selected, since it manages a trade-off between achievable performance and hardware requirements. While employing a binary detection of the received radiation is far simpler, obtaining the information regarding the power of said radiation could help to pinpoint the receiver. In Fig. \ref{fig:noise_vs_error}, the proposed system accuracy is plotted against multiple noise levels, for the original (floating-point) simulation data and for a binary version of the same data. For the assessed binary dataset, the average error rages from $6.2$ m (noiseless data) to $12.1$ m ($\sigma = 10$ dB), with a $95^{th}$ percentile error never exceeding $35.3$ m. As for the floating-point dataset, the average error rages from $4.3$ m to $13.6$ m, with a maximum $95^{th}$ percentile error of $40.4$ m. Therefore, from the obtained results, it is possible to conclude that using binary data is actually helpful for the CNN at higher noise regimes.

In Fig. \ref{fig:cdf}, the cumulative distribution function (CDF) for the binary dataset with moderate noise ($\sigma = 6$ dB) is shown. From the CDF, its is observable that the least accurate predictions are very inaccurate, explaining the significant difference between the median and the average errors ($6.1$ m and $10.6$ m, respectively). For this configuration, the predictions have a RMSE of $22.1$ m, which denotes superior performance when compared to the RMSE of $35$ m obtained in \cite{outdoor_fingerprint_35m} (which also considers a lower noise, with $\sigma=5$ dB). However, it is important to point out that the numerical simulations performed in \cite{outdoor_fingerprint_35m} do not contain NLOS positions.

\begin{figure}[t!]
  \centering
	\includegraphics[width=90mm, trim={4mm 5mm 2mm 5mm}]{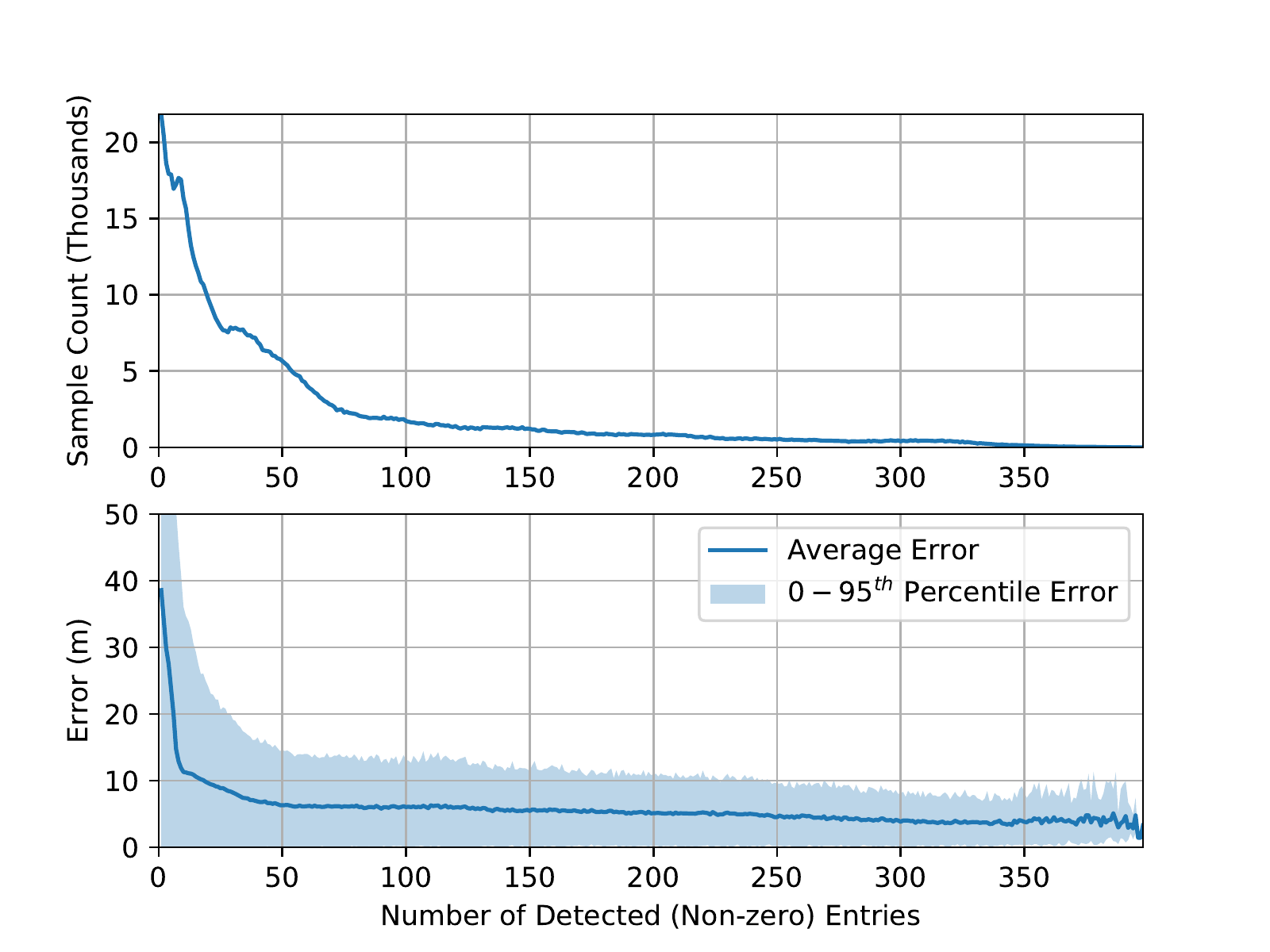}
	\caption{Prediction error and sample count per number of detected non-zero entries in a sample, considering binary data samples and a noise $\sigma$ of $6$ dB. As observed, when few entries are detected, the prediction error soars, explaining the least accurate predictions shown in Fig. \ref{fig:cdf}.}
	\label{fig:binned_error}
\end{figure}

To explain the inaccurate predictions, Fig. \ref{fig:binned_error} plots the error against the number of detected (non-zero) entries per sample, for the aforementioned dataset and noise parameters. The general trend is a lower average error with an increasing number of detected entries per sample. Furthermore, it is important to notice that when a sample contains fewer than $10$ non-zero entries, the average prediction error soars due to the lack of information in the received data, explaining the last percentiles of the CDF. Therefore, due to the mmWave propagation characteristics, the presence of multiple paths between the transmitter and the receiver enhances the information perceived by the proposed system, resulting in improved predictions. Furthermore, since additional sources of information improve the system performance, the presence of non-zero entries originating from other BSs should increase the prediction accuracy, making the proposed system fit for dense BS deployment schemes.

\begin{figure}[t!]
  \centering
	\includegraphics[width=90mm, trim={4mm 5mm 2mm 5mm}]{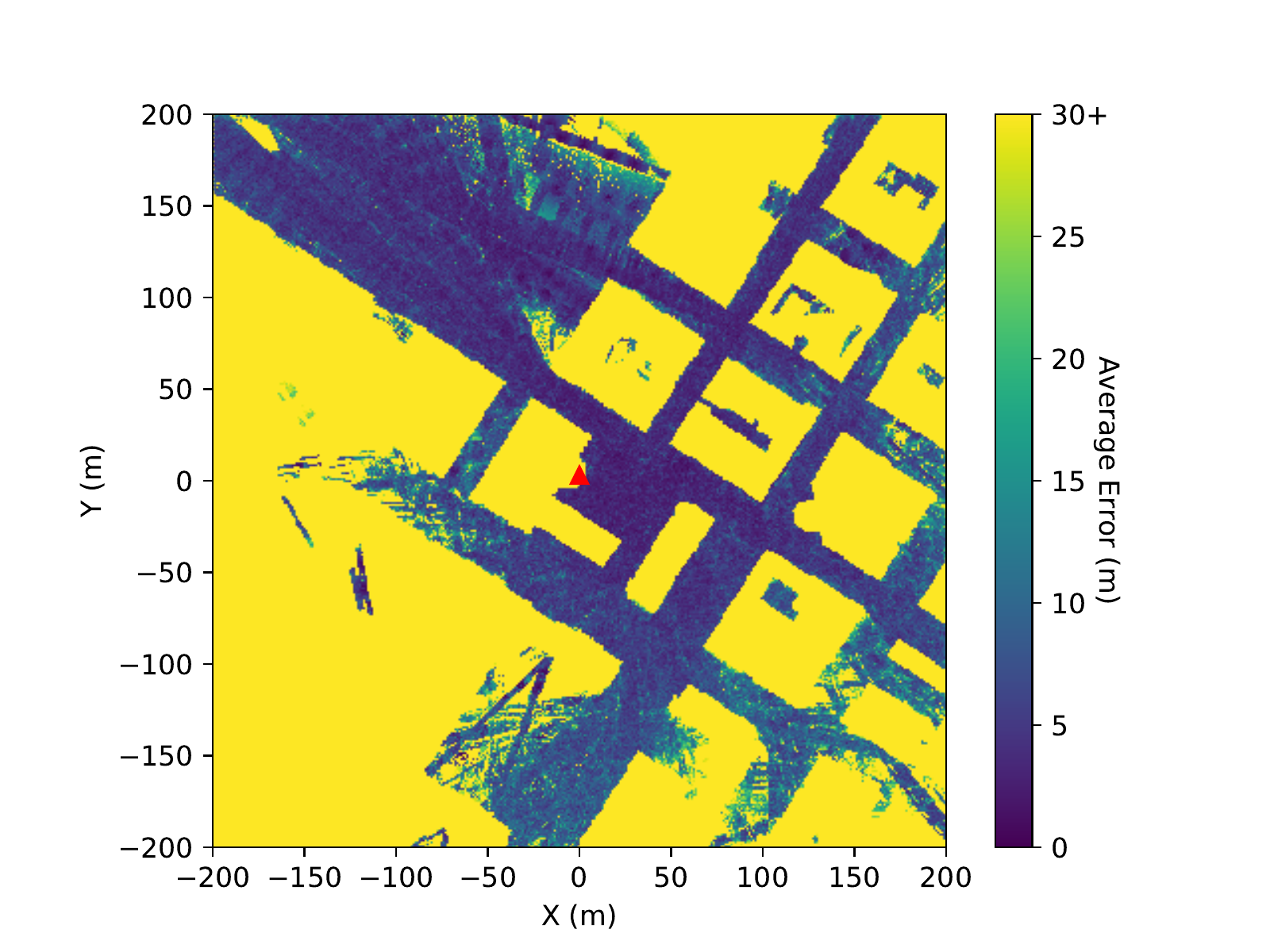}
	\caption{Average error per considered position, assuming binary data samples and a noise $\sigma$ of $6$ dB. Given that the transmitter is at the center of the image (red triangle), it is possible to verify that being in a NLOS position is not a constraint for the proposed system.}
	\label{fig:error_map}
\end{figure}

Finally, in Fig. \ref{fig:error_map} the average prediction error per position is shown, considering once again a binary dataset with a noise $\sigma$ of $6$ dB. Comparing these results to Fig. \ref{fig:nyu_sim}, it is possible to conclude that the system is always able to return a positioning prediction, as long as the mobile user is covered by the mmWave network. Moreover, it is interesting to notice that the prediction quality is not dictated by the received power: for instance, inside the building block at ($75$, $50$), the mobile receiver is expected to have a weak signal and a high-quality position estimate. This can be easily explained due to a unique data pattern, which is identified by the CNN. As a final key remark, it is clearly observable that being in a NLOS position is not a constraint to obtain an accurate positioning prediction.

\section{Conclusions}

Throughout this paper, a system that is able to predict a device geolocation through mmWave transmissions and CNNs is proposed. This system will be able to function whenever there is 5G coverage, and should require no additional hardware in devices that have mmWave reception capabilities. 

In the preliminary simulations performed, the average estimation error can drop below $10$ meters, in a scenario containing mostly NLOS positions, outperforming the existing algorithms for such positions. Therefore, by providing accurate estimates for NLOS positions, where other sub-meter-accuracy mmWave positioning algorithms struggle, the proposed system can be seen as a enabling component of the mmWave positioning techniques ecosystem, enhancing the end-user experience.

\bibliographystyle{IEEETran}
\bibliography{5g}

\end{document}